\documentclass[preprint,12pt]{elsarticle}
\usepackage{amssymb}
\usepackage{amssymb}
\usepackage{amsmath}
\allowdisplaybreaks[4]
\usepackage{lipsum,soul}
\usepackage{bm}
\usepackage{graphicx}
\usepackage{graphics}
\usepackage{amsmath}
\usepackage{array}
\usepackage[caption=false]{subfig}
\usepackage{xcolor}
\usepackage{subfig}
\usepackage{hyperref}
\usepackage{gensymb}
\usepackage[capitalise]{cleveref}
\usepackage{textcomp}

\usepackage{epstopdf}
\usepackage{cases}
\usepackage{amssymb}
\usepackage{multirow}
\usepackage{siunitx}
\usepackage{cases}
\usepackage{tabularx}
\usepackage{nomencl}
\usepackage{makecell}
\usepackage[top=1.2in, bottom=1.2in, left=1.1in, right=1.1in]{geometry}
\usepackage{tikz}
\usepackage{comment}
\usepackage{natbib}
\usepackage{lineno,hyperref}

\modulolinenumbers[1]
\usepackage{booktabs}

\begin{document}

\begin{frontmatter}

    \title{A new approach for the implementation of contact line motion based on the phase-filed lattice Boltzmann method}

    
    \author[inst1]{Long Ju}
    \ead{long.ju@kaust.edu.sa}
    
    \affiliation[inst1]{organization={Computational Transport Phenomena Laboratory (CTPL), King Abdullah University of Science and Technology (KAUST)},
                city={Thuwal},
                postcode={23955-6900}, 
                country={Kingdom of Saudi Arabia}}
    \author[inst3]{Zhaoli Guo}
    \ead{zlguo@hust.edu.cn}
    \author[inst2]{Bicheng Yan\corref{cor1}}
    \ead{bicheng.yan@kaust.edu.sa}
    \author[inst1]{Shuyu Sun\corref{cor1}}
    \ead{shuyu.sun@kaust.edu.sa}
    \cortext[cor1]{Corresponding author}
    
    \affiliation[inst3]{organization={Institute of Interdisciplinary Research for Mathematics and Applied Science, Huazhong University of Science and Technology},
                city={Wuhan},
                postcode={430074}, 
                country={China}}
    
    \affiliation[inst2]{organization={Energy Resource and Petroleum Engineering Program, Physical Science and Engineering Division, King Abdullah University of Science and Technology (KAUST)},
                city={Thuwal},
                postcode={23955-6900}, 
                country={Kingdom of Saudi Arabia}}
    
    \begin{abstract}
    This paper proposes a new strategy to implement the free-energy based wetting boundary condition within the phase-field lattice Boltzmann method. The greatest advantage of the proposed method is that the implementation of contact line motion can be significantly simplified while still maintaining good accuracy. For this purpose, the liquid-solid free energy is treated as a part of the chemical potential instead of the boundary condition, thus avoiding complicated interpolations with  irregular geometries. Several numerical testing cases including the droplet spreading processes on the idea flat, inclined and curved boundaries are conducted, and the results demonstrate that the proposed method has good ability and satisfactory accuracy to simulate contact line motions.
    \end{abstract}
    
    
    
    
    \end{frontmatter}
    
    
    \section{Introduction}
    \label{sec:sample1}
    Multiphase flows are frequently encountered in industrial operations and engineering applications, such as enhanced oil recovery~\cite{alvarado2010enhanced,yan2018multi}, geological carbon sequestration~\cite{zhang2014mechanisms,depaolo2013geochemistry,tariq2023spatial}, geothermal~\cite{yan2023robust} and underground hydrogen storage~\cite{huang2023compositional,song2021pore} as well. In these processes, the contact-line dynamics have long been of interest to the fluid research community, and intensive theoretical\cite{de1985wetting,cox1986dynamics1,cox1986dynamics2}, experimental\cite{zhang2008theoretical,wei2023experimental} and numerical studies\cite{mukherjee2007numerical,chaudhary2014freezing} have been performed. With significant advancements in computational capabilities, numerical modeling has emerged as an increasingly efficient approach. In spite of traditional numerical methods, such as level set method, volume-of-fluid method, the lattice Boltzmann equation (LBE) method rooted in kinetic theory, has developed as a powerful tool for simulating contact-line dynamics due to its innate kinetic nature, excellent adaptability for parallel computing, and ease in dealing with irregular boundaries\cite{guo2013lattice}.
    To date, numerous LBE models for multiphase flows have been developed based on diverse physical perspectives including the color-gradient model\cite{liu2016lattice,ba2016multiple,akai2018wetting}, the pseudopotential model\cite{chen2014critical,li2013lattice}, the free-energy model\cite{li2007symmetric,soomro2023fugacity,guo2021well}, and the phase-field-based model~\cite{fakhari2010phase,shu2013direct,yue2022improved}. Compared with other models, the phase-field LBE model has attracted much attention, due to its simplicity and accuracy. Since the total mass variation is managed by the interface tracking equation, implementation of the wetting boundary conditions for the phase distribution function in the phase-field LBE model becomes more straightforward.
    
    Within the LBE method community, the wettability of the solid boundaries is typically characterized by contact angles. Wetting boundary schemes are required to dictate the phase distribution at boundary nodes and achieve desired contact angles. According to a recent work proposed by Zhang et, al.~\cite{zhang2023simplified}, the mostly used wetting boundary treatments for phase-field LBE model can be categorized into three types. 
    The first approach, developed by Martys and Chen~\cite{martys1996simulation}, and later integrated into the phase-field LB model by Iwahara et al~\cite{iwahara2003liquid}, utilizes an artificial solid density to simulate fluid-solid interactions, positing solids as a two-phase fluid mixture. This straightforward and prevalent wetting boundary scheme is particularly apt for curved boundaries and numerous successful porous-media flow studies have been conducted based on it. However, the most disadvantage of this type of wetting boundary treatment is that contact angle is not input parameter, but have an implicit relationship with the solid density, which needs the extra pre-numerical simulations~\cite{zhang2023simplified}.
    The second type of wetting boundary treatment developed by Ding and Spelt~\cite{ding2007wetting} is from the viewpoint of the geometrical relation. 
    Compared with the first type of the boundary treatment, this boundary scheme can be mathematically proven that the numerically imposed contact angle can be guaranteed to be the exact prescribed value explicitly. Many contributions have been made to implement this method based on phase-field LB models and it has been successfully applied to the LBE method simulations of fluid-solid wetting phenomena~\cite{wang2013scheme,liu2015diffuse,huang2018alternative,huang2022simplified}.
    The last type of wetting treatment was developed by Briant~\cite{briant2002lattice,briant2004lattice}. Based on the surface-energy method, and Lee and Liu~\cite{lee2008wall} extended this method into the phase-field LBE model. In this scheme, the driven force of the motion of contact line is the surface energy, which is regarded as part of the free energy of the system. The surface energy related to the gradient of the order parameter $\phi$ is considered as the wetting boundary condition, which can be expressed as $\kappa \bm{n}_w\cdot \bm{\nabla}\phi=d\psi_s/d\phi$, where $\bm{n}_w$ is the normal vector pointing from solid to fluid, $\psi_s$ represents the surface energy and $\kappa$ is the positive free-energy coefficient. 
    Compared to the above two, this type of wetting boundary treatment can intuitively and accurately simulate a given contact angle with a solid physical foundation, so the present work primarily focuses on this kind of boundary treatments.
    
    In the original models, surface energy was defined as a linear function of the order parameter, which inadvertently introduced an undesired mass layer~\cite{2009WALL}. To address this, recent adaptations have explored alternative functional representations for surface energy, including sine and cubic functions~\cite{qian2003molecular,villanueva2006some,khatavkar2007capillary}. Taking the cubic form as an example, the wetting boundary condition can be expressed as $\bm{n}_w\cdot \bm{\nabla}\phi=-\sqrt{2\beta/\kappa}\cos\theta(\phi-\phi^2)$, where $\beta$ is a physical parameter determined by the given interface thickness and surface tension.
    In the calculation, the above boundary condition is actually adopted to determine the order parameters on ghost solid node. Obviously, it is relatively easy to handle this for a flat boundary, where the normal vector $\bm{n}_w$ points to a given lattice node. The order parameter on the ghost cell can be obtained by solving the aforementioned quadratic equation, with the gradient term directly determined through interpolation.
    While for the the curved boundaries, the calculation of the ghost solid order parameters is very complicated.
    To simplify the implementation, Connington and Lee~\cite{CONNINGTON2013601,CONNINGTON2015453} were effectively assuming that the normal vector is in the direction along with the lattice link, pointing away from the solid.
    This approach offers a degree of simplification in boundary handling, but it comes at the expense of computational accuracy.
    Apart from that, 
    Fakhari and Bolster~\cite{FAKHARI2017620,FAKHARI2018119} adopted a biquadratic interpolation to determine the order parameter along the normal vector. Given its implicit nature, this method might necessitate iterative solutions. Although the authors introduced a simplified version in their studies, but the linear or quadratic interpolations are still required, and an additional directional judgments are introduced in the calculation, which complicates its implementation.
    To the best of our knowledge, there remains a trade-off between implementation accuracy and complexity when dealing with the free-energy wetting boundary in the phase-field LB model. A boundary treatment with a clear physical basis, simple implementation, and good accuracy is still needed.
    
    To achieve this objective, a simplified implementation of wetting boundary condition for Cahn-Hilliard (CH) based phase-filed LBE method is proposed in this work. 
    The remainder of the present paper is organized as follows. In~\cref{sec2}, the details of implementation for the wetting boundary condition are introduced after given the governing equations and LBE method for two-phase flow.~\cref{sec3} provides the numerical validation to test the performance of the proposed boundary treatment. Finally, a summary is given in~\cref{sec4}.
    
    \section{Mathematical method}\label{sec2}
    \subsection{Governing equations and lattice Boltzmann method for two-phase flow}
    \subsubsection{Governing equations}
    The two-phase incompressible fluid flows can be governed by the Navier-Stokes equations and the CH equation, which can be expressed as~\cite{zhang2019fractional},
    \begin{subequations}
        \begin{equation}
            \bm{\nabla}\cdot \bm{u}=0,
        \end{equation}
        \begin{equation}
            \partial_t (\rho \bm{u})+\bm{\nabla}\cdot (\rho \bm{uu})=-\bm{\nabla}p+\bm{\nabla}\cdot \left[\rho\nu(\bm{\nabla\bm{u}+\nabla\bm{u}^{T}})\right]+\bm{F}_s+\bm{F}_b,
        \end{equation}
        \begin{equation}
            \partial_t \phi+\bm{u}\cdot\bm{\nabla}\phi=\bm{\nabla}\cdot(M\bm{\nabla}\mu_{\phi}),
        \end{equation}
    \end{subequations}
    where $\bm{u}$ is the velocity, $\rho$ is the density, $p$ is the pressure and $\nu$ is the kinematic viscosity. $\bm{F}_s$ donates the surface tension force, which is chosen as $\bm{F}_s=\mu_{\phi}\bm{\nabla}\phi$, with $\mu_{\phi}$ being the chemical potential. $\bm{F}_b$ is the body force. $\phi$ represents the order parameter, which is used to distinguish the different phases. In this work, the order parameter is set as 1 and 0 for liquid and vapor phases, respectively, with a diffuse phase interface from 0 to 1. $M$ represents the mobility.
    
    In a two-phase system, the density and viscosity is no longer homogeneous as it exhibits a discontinuity at the liquid-gas interface, which are all assumed to be a linear function of the order parameter here~\cite{liang2018phase}, 
    \begin{equation}
        \rho=\phi(\rho_1-\rho_0)+\rho_0, \qquad \nu=\phi(\nu_1-\nu_0)+\nu_0,
    \end{equation}
    \subsubsection{Lattice Boltzmann model for incompressible fluid flow}
    In the LBE method, the space is discretized into regular lattices, and all particle distribution functions (PDF) are assumed to move with a series of discrete velocities on the nodes. In the standard LBE mdoel, the evolution of these PDF can be described by~\cite{guo2013lattice}
    \begin{equation}\label{BGK-NS}
    f_i(\bm{x}+\bm{c}_i\delta_t, t+\delta_t)-f_i(\bm{x},t)=-\frac{1}{\tau_f}\left[f_i(\bm{x},t)-f_i^{eq}(\bm{x},t)\right]+\delta_t {F}_i(\bm{x},t),
    \end{equation}
    where $f_i(\bm{x}, t)$ is the PDF at position $\bm{x}$ and time $t$. $\bm{c}_i$ is the discrete velocity. In two dimensions (2D), the most popular D2Q9 (two-dimension-nine-velocity) is adopted here, and $\bm{c}_i$ is defined as
    \begin{equation}
    \bm{c}_i=
    \begin{cases}
    (0,0)c,& i=0\\
    (\cos[(i-2)\pi/2],\text{sin}[(i-2)\pi/2])c,& i=1\sim4,\\
    \sqrt{2}(\cos[(i-5)\pi/2+\pi/4],\text{sin}[(i-5)\pi/2+\pi/4])c,& i=5\sim8.\\
    \end{cases}
    \label{eq:ci}
    \end{equation}
    where $c=\delta_x/\delta_t$ is the lattice speed with $\delta_x$ and $\delta_t$ being the lattice spacing and time step, respectively. In three dimensions, the D3Q19 (three-dimension-nineteen-velocity) model is used, in which the discrete velocity can be expressed as
    \begin{equation}
    \bm{c}_i=
    \begin{cases}
    c(0,0,0),& i=0\\
    c(\pm{1},0,0), c(0,\pm{1},0),c(0,0,\pm{1}) & i=1\sim6,\\
    c(\pm{1},\pm{1},0), c(\pm{1},0,\pm{1}),c(0,\pm{1},\pm{1}),& i=7\sim18.\\
    \end{cases}
    \label{eq:3ci}
    \end{equation}
    $\tau_f$ in~\cref{BGK-NS} is the viscosity-related relation time. $f_i^{eq}$ is the equilibrium distribution function, which can be written as
    \begin{equation}\label{feq_i}
    f_i^{eq}=
    \begin{cases}
        \frac{p}{c_s^2}(\omega_i-1)+\rho {s}_i(\bm{u}),& i=0,\\
        \frac{p}{c_s^2}\omega_i+\rho {s}_i(\bm{u}),& i\neq 0,
    \end{cases}
    \end{equation}
    with ${s}_i(\bm{u})$ being written as
    \begin{equation}\label{Siu}
    {s}_i(\bm{u})=\omega_i\left[\frac{\bm{c}_i\cdot \bm{u}}{c_s^2}+\frac{(\bm{c}_i\cdot \bm{u})^2}{2c_s^4}-\frac{\bm{u}\cdot\bm{u}}{2c_s^2}\right],
    \end{equation}
    where $\omega_i$ is the weighting coefficient and $c_s$ is the sound speed, which are defined as 
    \begin{subequations}
    \begin{equation}
        \omega_0=\frac{4}{9}, \omega_{1-4}=\frac{1}{9}, \omega_{5-8}=\frac{1}{36}, c_s^2=\frac{c^2}{3},\quad \text{for D2Q9},
    \end{equation}
    \begin{equation}
         \omega_0=\frac{1}{3}, \omega_{1-6}=\frac{1}{18}, \omega_{7-18}=\frac{1}{36}, c_s^2=\frac{c^2}{3},\quad \text{for D3Q19},
    \end{equation} 
    \end{subequations}
    $F_i(\bm{x},t)$ in ~\cref{BGK-NS} symbolizes the force distribution function, which is elaborately designed as~\cite{guo2002discrete}
    \begin{equation}\label{GGi}
    F_i=(1-\frac{1}{2\tau_g})\omega_i\left[\bm{u}\cdot \bm{\nabla}\rho+\frac{\bm{c}_i\cdot \bm{F}}{c_s^2}+\frac{(\bm{u\nabla}\rho:(\bm{c}_i\bm{c}_i-c_s^2\bm{I})}{c_s^2}\right],
    \end{equation}
    where $\bm{F}=\bm{F}_s+\bm{F}_b$ is the total force.
    The fluid pressure and velocity in the present model can be calculated as
    \begin{subequations}
    \begin{equation}
            p=\frac{c_s^2}{(1-\omega_0)}\left[\sum_{i\neq 0}f_i+0.5\delta_t\bm{u}\cdot \bm{\nabla}\rho+\rho s_0(u)\right],
        \end{equation}
        \begin{equation}
            \rho\bm{u}=\sum_i\bm{c}_if_i+0.5\delta_t\bm{F},
        \end{equation}
    \end{subequations}
    Based on the Chapman-Enskog analysis~\cite{guo2013lattice}, the NS equations can be recovered from~\cref{BGK-NS} with the fluid kinematic viscosity determining by
    \begin{equation}
        \nu=c_s^2(\tau_f-0.5)\delta_t.
    \end{equation}
    
    \subsubsection{Lattice Boltzmann model for phase interface capture}
    
    For the phase interface capture, the well-balanced LBE model is adopted here~\cite{2311.10827}, in which the LB evolution equation with the BGK collision operator for the CH equation is expressed as,
    \begin{equation}\label{BGK}
    g_i(\bm{x}+\bm{c}_i\delta_t, t+\delta_t)-g_i(\bm{x},t)=-\frac{1}{\tau_g}\left[g_i(\bm{x},t)-g_i^{eq}(\bm{x},t)\right]+\delta_t {G}_i(\bm{x},t)+\frac{1}{2}\delta_t^2\partial_t G_i(\bm{x},t),
    \end{equation}
    with the equilibrium distribution function $g_i^{eq}$ being defined as
    \begin{equation}\label{H_i}
    g_i^{eq}=
    \begin{cases}
        \phi-(1-\omega_0)\alpha \mu_{\phi},& i=0,\\
        \omega_i\alpha \mu_{\phi},& i\neq 0,
    \end{cases}
    \end{equation}
    where $\alpha$ is an adjusted parameter. The source term $G_i$ is defined as
    \begin{equation}\label{F_i}
    G_i=\omega_i(\bm{u}\cdot \bm{\nabla} \phi)\left[-1+\frac{\bm{I}:(\bm{c}_i\bm{c}_i-c_s^2\bm{I})}{2c_s^2}\right],
    \end{equation}
    D2Q9 and D3Q7 (three-dimension-seven-velocity) are adopted here for two and three dimensional calculations, respectively. In D3Q7 model, the discrete velocity $\bm{c}_i$ is defined as
    \begin{equation}
    \bm{c}_i=\begin{pmatrix} 
    0 & 1 & -1 & 0 & 0 & 0 & 0\\ 
    0 & 0 &  0 & 1 & -1 & 0 & 0\\ 
    0 & 0 &  0 & 0 & 0 & 1 & -1
    \end{pmatrix}.
    \end{equation}
    The weighting coefficient and sound speed are defined as
    \begin{equation}
        \omega_0=\frac{1}{4}, \omega_{1-6}=\frac{1}{8}, c_s^2=\frac{c^2}{4}.
    \end{equation}
    The order parameter in the present model can be computed by 
    \begin{equation}\label{order-parameter}
    \phi=\sum_ig_i.
    \end{equation}
    Applying the CE analysis to the LB equation~\ref{BGK}, the CH equation can be exactly recovered without any additional assumption, and the relation between the mobility $M$ and the relation time $\tau_g$ can be expressed as
    \begin{equation}\label{mobility}
    M_{\phi}=c_s^2\alpha(\tau_g-0.5)\delta_t.
    \end{equation}
    
    To achieve precise numerical computations, it's essential to employ suitable difference schemes when discretizing the model's derivative terms. For simplicity, the gradient term can be determined using a second-order isotropic central scheme~\cite{liang2014phase,liang2018phase},
    \begin{equation}\label{gradient}
    \bm{\nabla}\Gamma(\bm{x})=\sum_{i\neq 0}\frac{\omega_i\bm{c}_i\Gamma(\bm{x}+\bm{c}_i\delta_t)}{c_s^2\delta_t},
    \end{equation}
    and the Laplace operator can be calculated by
    \begin{equation}\label{Laplace}
    {\nabla}^2\Gamma(\bm{x})=\sum_{i\neq 0}\frac{2\omega_i[\Gamma(\bm{x}+\bm{c}_i\delta_t)-\Gamma(\bm{x})]}{c_s^2\delta^2_t},
    \end{equation}
    where $\Gamma$ is the any physical variable.
    
    \subsection{Wetting boundary condition}
    \subsubsection{The construction of the wetting boundary condition}
    When two-phase fluids come into contact with a solid substrate, the substrate's wettability significantly impacts fluid interface dynamics. Therefore, it is essential to establish a wetting boundary condition that incorporates the contact angle between the phase interface and the solid surface. In this subsection, we will give the details on the construction of the wetting boundary conditions, and propose a simplified implementation for it based on the above phase-filed LBE model.
    
    First, we denote a multiphase flow domain by $\Omega$ and its solid boundary by $\partial\Omega$, then the total free energy of this system can be expressed as~\cite{moldover1980interface,liang2019lattice,de1985wetting}
    \begin{equation}\label{free-energy}
        \mathcal{F}=\int_{\Omega} \Psi(\phi, \nabla \phi) d\Omega=\int_{\Omega} \left[\psi(\phi)+\frac{\kappa}{2}|\nabla \phi|^2\right]d\Omega+\int_{\partial_{\Omega}} \psi_s(\phi)ds,
    \end{equation}
    where $\psi(\phi)$ is the bulk free-energy density and is chosen to have a double-well form~\cite{liang2018phase} in this work,
    \begin{equation}\label{bulk_energy}
    \psi(\phi)=\beta\phi^2(\phi-1)^2, 
    \end{equation}
    $\frac{\kappa}{2}|\nabla \phi|^2$ accounts for the phase-interface free energy density with $\kappa$ being a positive free-energy coefficient, and $\psi_s$ is the free-energy density on the fluid-solid boundary. Applying the variational operator to~\cref{free-energy}, we can obtain
    \begin{equation}
        \delta\mathcal{F}=\int_{\Omega} \left[\frac{\partial\psi}{\partial \phi}\delta \phi+\kappa \bm{\nabla}\phi\cdot \delta(\bm{\nabla}\phi)\right] d\Omega+\int_{\partial_{\Omega}} \frac{\partial\psi_s}{\partial \phi}\delta \phi ds,
    \end{equation}
    which can be further written as follows by using the Gauss integral theorem~\cite{liang2019lattice},
    \begin{equation}\label{FF}
        \delta\mathcal{F}=\int_{\Omega} \left[\frac{\partial\psi}{\partial \phi}-\kappa {\nabla}^2\phi\right] \delta \phi d\Omega +\int_{\partial_{\Omega}} \left[-\kappa\bm{n}_w\cdot \bm{\nabla}\phi+\frac{\partial\psi_s}{\partial \phi}\right]\delta \phi ds.
    \end{equation}
    Obviously, how to specify the wall free-energy density $\psi_s$ is important. Similar with the exiting studies, the cubic wall free energy is adopted with the interactions between solid and bulk phases neglected, and only the interaction at the three-phase junction is considered~\cite{martys1996simulation,khatavkar2007capillary},
    \begin{equation}\label{cubic}
        \psi_s=\frac{b_1}{2}\phi^2-\frac{b_1}{3}\phi^3,
    \end{equation}
    then $\partial\psi_s/\partial\phi=b_1(\phi-\phi^2)$, and $b_1$ is still need to be specified. Referring to reference~\cite{liang2019lattice}, another constraint condition in the bulk region can be derived according to~\cref{FF},
    \begin{equation}\label{sup}
        \frac{d\psi_s}{d\phi}=\pm\sqrt{2\kappa \psi}.
    \end{equation}
    Combining~\cref{bulk_energy,cubic,sup}, we can found that~\cref{sup} has two stable solutions of $\phi_{s1}=0$ and $\phi_{s2}=1$. Subsequently, the surface tensions of the gas-solid and liquid-solid phases can be represented as~\cite{yan2007lattice}
    \begin{subequations}
        \begin{equation}
            \sigma_{sg}=\frac{b_1}{2}\phi_{s1}^2-\frac{b_1}{3}\phi_{s1}^3+\int_0^{\phi_{s1}}\sqrt{2\kappa\psi}d\phi=0,
        \end{equation}
        \begin{equation}
            \sigma_{sl}=\frac{b_1}{2}\phi_{s2}^2-\frac{b_1}{3}\phi_{s2}^3+\int_1^{\phi_{s2}}\sqrt{2\kappa\psi}d\phi=\frac{b_1}{6}.
        \end{equation}
    \end{subequations}
    Then, for the two-phase fluids on the chemically homogeneous wall, the contact angle can be determined based on the Young’s equation~\cite{young1805iii},
    \begin{equation}\label{b1}
        \cos{\theta}=\frac{\sigma_{sg}-\sigma_{sl}}{\sigma}=-\frac{b_1}{\sqrt{2\kappa \beta}}.
    \end{equation}
    According to~\cref{cubic,b1}, the free-energy functional can be written as
    \begin{align}\label{rs}
        \delta\mathcal{F}=&\int_{\Omega} \left[\frac{\partial\psi}{\partial \phi}-\kappa {\nabla}^2\phi\right] \delta \phi d\Omega \\\nonumber
        &+\int_{\partial_{\Omega}} \left[-\kappa\bm{n}_w\cdot \bm{\nabla}\phi-\sqrt{2\kappa\beta}\cos\theta(\phi-\phi^2)\right]\delta \phi ds,
    \end{align}
    where $\bm{n}_w$ is the normal vector pointing from solid to the fluid.
    In fact, by introducing the effective surface area i.e., $a_v$, the surface integral in the above equation can be transformed into a volume integral~\cite{li2009solving}, which can be expressed as
    \begin{align}\label{rs2}
        \delta\mathcal{F}=&\int_{\Omega} \left[\frac{\partial\psi}{\partial \phi}-\kappa {\nabla}^2\phi
        -a_v\kappa\bm{n}_w\cdot \bm{\nabla}\phi-a_v\sqrt{2\kappa\beta}\cos\theta(\phi-\phi^2)\right]\delta \phi d\Omega,
    \end{align}
    The variation of free energy $\mathcal{F}$ with regard to order parameter is referred to as the chemical potential~\cite{guo2021well},
    \begin{align}\label{chemical_potencial11}
    \mathcal{\mu}_{\phi}&=\frac{d \psi}{d{\phi}}-\kappa\nabla^2{\phi}-a_v\left[\kappa\bm{n}_w\cdot \bm{\nabla}{\phi}+\sqrt{2\kappa\beta}\cos\theta({\phi}-{\phi}^2)\right],
    \end{align}
    The first two terms in the right hand side of the~\cref{chemical_potencial11} donate the chemical potentials in the bulk area and the phase interface. The third term represents that on the domain boundary $\partial \Omega$. 
    Actually, the~\cref{chemical_potencial11} is usually implemented coupled with the boundary treatments,
    \begin{equation}\label{cubicc}
        \bm{n}_w\cdot \bm{\nabla}\phi=\chi,
    \end{equation}
    with~\cref{chemical_potencial11} being expressed as
    \begin{align}\label{chemical_potencial112}
    \mathcal{\mu}_{\phi}&=\frac{d \psi}{d{\phi}}-\kappa\nabla^2{\phi}-a_v\left[\kappa\chi+\sqrt{2\kappa\beta}\cos\theta({\phi}-{\phi}^2)\right],
    \end{align}
    where any expressions of $\chi$ can lead to the preset wetting conditions, which has been proved numerically in appendix A. If $\chi=-\sqrt{{2\beta}/{\kappa}}\cos\theta(\phi-\phi^2)$ is chosen, ~\cref{cubicc,chemical_potencial112} then become  
    usually used cubic wetting boundary condition.
    As discussed in the introduction, it is complicated to implement the above boundary condition for the boundary with irregular geometry. To simplify the boundary treatment,we take $\chi=0$ to switch~\cref{cubicc} from Robin boundary condition to the Neumann's type here,
    \begin{equation}\label{nowet}
        \bm{n}_w\cdot\bm{\nabla}\phi=0,
    \end{equation}
    obviously, compared with the typical cubic wetting boundary condition, the implementation of~\cref{nowet} is much simpler. The chemical potential in a control volume then can be expressed as
    \begin{align}\label{chemical_potencial2}
    \mathcal{\mu}_{\phi}
    =4\beta{\phi}({\phi}-1)({\phi}-0.5)-\kappa \nabla^2 {\phi}-a_v\sqrt{2\kappa\beta}\cos\theta({\phi}-{\phi}^2),
    \end{align}
    The combination of boundary condition~\cref{nowet} and chemical potential~\cref{chemical_potencial2} can describe the contact line motion of two-phase fluids.
    
    \subsubsection{The implementation of the boundary conditions}
\begin{figure}[ht]
         \centering
        {\includegraphics[width=0.8\textwidth]{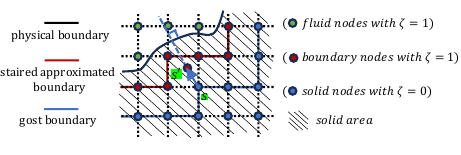}}~~
         \caption{Schematic illustration for the lattice nodes around the physical boundary.}
         \label{boundary}
    \end{figure}

    ~\cref{boundary} shows a two-dimensional schematic illustration for some lattice nodes near the boundary. A mark symbol $\zeta$ is introduced to distinguish the fluid ($\zeta=1$)and solid ($\zeta=0$) nodes, and it is noted that the boundary nodes is marked as the same as the fluid node as $\zeta=1$, due to the no-slip and no-flux boundary conditions are all implemented based on the modified bounce-back scheme in the present study, where the boundary nodes also participate in the collision and streaming processes.  
    
    For the no-slip and no-flux boundary conditions, the unknown PDFs $f_i(\bm{x}_b)$ and $g_i(\bm{x}_b)$ can be calculated by the modified bounce-back scheme after the streaming step,
    \begin{equation}
        f_i(\bm{x}_b)=f_{i'}(\bm{x}_b), \qquad g_i(\bm{x}_b)=g_{i'}(\bm{x}_b),
    \end{equation}
    where $i'$ is the opposite direction of $i$. To capture the wetting phenomena, ~\cref{nowet} should be used to determine the order parameter $\phi$ on the ghost lattice nodes. As shown in~\cref{boundary}, $\phi(\bm{x}_g)$ should be equal to $\phi(\bm{x}_{g'})$, which is approximated by the average value of its surrounding nodes~\cite{li2019implementation,zhang2023simplified},
    \begin{equation}
    \phi(\bm{x}_s)=\phi(\bm{x}_{s'})\approx\frac{\sum_i^b\zeta(\bm{x}_g+\bm{c}'_i\delta_t)\phi(\bm{x}_g+\bm{c}'_i\delta_t)}{\sum_i^b\zeta(\bm{x}_g+\bm{c}'_i\delta_t)},
    \end{equation}
    where $b$ is the total linked directions of a ghost node. For 2D simulations, $b=9$ and $\bm{c}'_i=\bm{c}_i$ in~\cref{eq:ci}, but for 3D cases, $b=27$ and $\bm{c}'_i$ can be expressed as
    \begin{equation}
    \bm{c}'=\begin{pmatrix} 
    \bm{M} & \bm{N} 
    \end{pmatrix},
    \end{equation}
    where $\bm{M}$ is the discrete velocity $\bm{c}_i$ in D3Q19 model in~\cref{eq:3ci}, and $\bm{N}$ donates its supplement,
    \begin{equation}
    \bm{N}=c\begin{pmatrix} 
    1 & -1 & 1 & -1 & 1 & -1 & 1 & -1\\ 
    1 & 1 &  -1 & -1 & 1 & 1 & -1 & -1\\ 
    1 & 1 &  1 & 1 & -1 & -1 & -1 & -1
    \end{pmatrix}.
    \end{equation}
    \subsubsection{Approximation of the effective surface area $a_v$} 
    In our proposed wetting boundary treatment, the effective surface area $a_v$ is a crucial parameter, which is needed to be calculate with great care. For a flat boundary, the effective surface area is precisely equal to $1/\delta_x$. However, it is difficult to obtain its precise value for a solid boundary with irregular geometries, and here we approximate it by~\cite{whitaker1998method,soulaine2017mineral}
    \begin{equation}\label{avvv}
        a_v\approx|\bm{\nabla}\epsilon|,
    \end{equation}
    It should be noted that the above equation starts from the volume averaging theorem, which links the porosity gradient to the average surface normal within a control volume, i,e.,
    $-\bm{\nabla}\epsilon=(1/V)\int_{\partial_V}\bm{n}_wdA$. 
    $\epsilon$ in~\cref{avvv} is the volume fraction of void space within a control volume, which is defined as 
    \begin{equation}\label{H_i}
    \epsilon=
    \begin{cases}
        1,& \text{fluid nodes},\\
        0,& \text{boundary nodes and solid nodes}.
    \end{cases}
    \end{equation}
    \begin{figure}[ht]
         \centering
        {\includegraphics[width=0.9\textwidth]{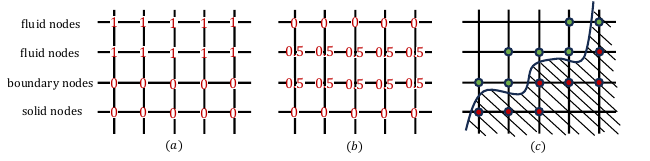}}~~
         \caption{Schematic illustration for (a) volume fraction $\epsilon$, (b) the calculated effective surface area $a_v/dx$ and (c) lattice nodes involved in calculating the effective surface area of a curved boundary, and red dots represent boundary nodes, and the green ones denotes fluid nodes.}
         \label{av}
    \end{figure}
    It should be noted that $\epsilon$ at boundary nodes is set as zero, which is different with $\zeta$. The gradient of $\epsilon$ can be calculated by~\cref{gradient}. However, there is a sharp transition between the fluid region and the solid region, As shown in~\cref{av}(a) and (b), ~\cref{gradient} actually extends the wetting boundary effects to the adjacent fluid nodes. Actually, this two-layer's structure can ensure better accuracy when dealing with curved boundaries with the stair-step approximation. As shown in~\cref{av}(c), the lattice nodes in the solid phase (marked as red dot) usually used to approximate the curved boundary, obviously, this approximation will extend the actual physical boundary towards the solid phase region which could introduce significant deviations. However, the structure of the double-layer surface chemical potential could naturally involve nearby fluid points (marked as green dot), and lattice nodes on both sides of the physical boundary are used to calculate the surface chemical potential, ensuring that the calculated chemical potential is in the vicinity of the actual physical surface. The accuracy of this treatment would been validated in subsequent testing cases. 
    
    \section{Numerical test and discussions}\label{sec3}
    In this section, several benchmark examples including the droplet spreading on both flat, inclined and curved ideal walls, are going to be performed to validate the accuracy of our proposed wetting boundary treatments in both 2D and 3D phase-field LB simulations. D2Q9 and D3Q19 models are adopted in 2D and 3D simulation cases, respectively. 
    \subsection{Droplet spreading on the flat ideal wall}\label{sec3.1}
\begin{figure}[ht]
         \centering
        {\includegraphics[width=0.6\textwidth]{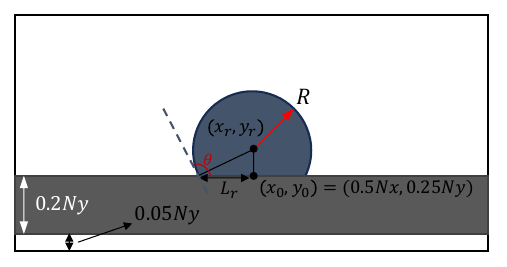}}~~
         \caption{Schematic illustration for the wetting of a droplet on an idea flat surface.}
         \label{flatshiyi}
    \end{figure}

\begin{figure}[ht]
         \centering
        {\includegraphics[width=0.8\textwidth]{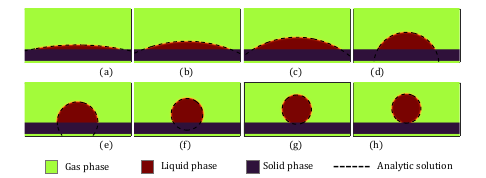}}~~
         \caption{The predicted droplet equilibrium shapes in 2D using the present boundary treatment with wide range of prescribed contact angles, (a) $\theta=10\degree$, (b) $\theta=20\degree$, (c) $\theta=30\degree$, (d) $\theta=60\degree$, (e) $\theta=90\degree$, (f) $\theta=120\degree$, (g) $\theta=150\degree$, (h) $\theta=160\degree$.}
         \label{2Dflat}
    \end{figure}

    A fundamental two-phase droplet spreading problem on an ideal wall is initially employed to validate the capability of the established numerical approach in predicting the wide range of contact angles. The simulations are performed in $Nx \times Ny = 256 \times 128$ rectangular domain for 2D simulations. A semicircular droplet with the radius $R = 50$ is initially deposited on the flat solid wall. The thickness of the solid plate is $0.25Ny$, which is $0.05Ny$ from the bottom (as displayed in~\cref{flatshiyi}). It noted that for very low contact angles of $\theta = 20\degree$ and $10\degree$, the steady droplet has exceeded the grid space, and then we have decreased the radius of the droplet to $R=40$ and $30$, respectively.
    To match this setup, the initial distribution profile of the order parameter is given by~\cite{liu2022diffuse}
    \begin{equation}\label{initialphi}
        \phi(x,y)=0.5+0.5\tanh \frac{2[R-\sqrt{(x-x_0)^2+(y-y_0)^2}]}{W}.
    \end{equation}
    The analytical solution of the order parameter for the droplet at the equilibrium state can be expressed as 
    \begin{equation}
        \phi_{r}=0.5+0.5\tanh \frac{2[R_r-\sqrt{(x-x_r)^2+(y-y_r)^2}]}{W},
    \end{equation}
    where $R_r=L_r/\sin(\theta)$, $x_r=x_0$ and $y_r=y_0-R_r\cos\theta$.
    In the simulation, some physical parameters are set as $\rho_l=10.0$, $\rho_g=1.0$, $\nu_l=\nu_g=0.1$, $\sigma=0.005$, $M=0.01$ and $W=4$. The periodic boundary condition is applied in all surrounding boundaries, and the wetting boundary treatments are adopted for the fluid-solid interface.
    \cref{2Dflat} illustrates the droplet equilibrium shapes predicted by the LBE method, incorporating the proposed surface energy wetting boundary treatment across a broad spectrum of specified contact angles. As evident from~\cref{2Dflat}, the droplet can assume various stable configurations on the substrate, which are significantly influenced by the specified contact angle, and the numerical results are all agree well with the analytical solutions.
    \begin{figure}[ht]
         \centering
        {\includegraphics[width=0.7\textwidth]{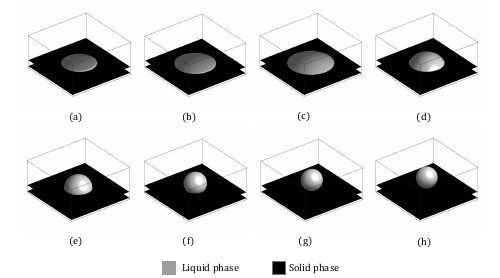}}~~
         \caption{The predicted droplet equilibrium shapes in 3D using the present boundary treatment with wide range of prescribed contact angles, (a) $\theta=10\degree$, (b) $\theta=20\degree$, (c) $\theta=30\degree$, (d) $\theta=60\degree$, (e) $\theta=90\degree$, (f) $\theta=120\degree$, (g) $\theta=150\degree$, (h) $\theta=160\degree$.}
         \label{3Dflat}
    \end{figure}
    \begin{table}[ht]
        \centering
        \caption{Simulation results of the proposed wetting schemes in predicting wide range of contact angles}
        \begin{tabular}{ccccc}
        \toprule[1.5pt]
            \makebox[0.2\textwidth][c]{Contact angle (\degree)} & \makebox[0.14\textwidth][c]{\makecell{2D results (\degree)}} & \makebox[0.14\textwidth][c]{\makecell{3D results (\degree)}} & \makebox[0.14\textwidth][c]{\makecell{2D errors (\degree)}} & \makebox[0.14\textwidth][c]{\makecell{3D errors (\degree)}} \\
            \midrule[1pt]
            10 &  10.2&  12.3&  0.2& 2.3\\
            20 &  19.3&  19.7&  0.7&0.3\\
            30 &  29.6& 30.2 & 0.4 & 0.2\\
            40 & 39.2& 39.7 & 0.8 &0.3\\
            60 & 59.3 &  59.5& 0.7 & 0.5\\
            90 &  89.5& 89.9 & 0.5 &0.1 \\
            120 & 119.9 & 120.7 & 0.1 & 0.7\\
            140 & 140.2 & 142.8 & 0.2 & 2.8\\
            150 & 150.1 & 156.3 & 0.2 & 6.3\\
            160& 159.7 & 180 & 0.3 & -\\
            170& 168.3 & 180 & 1.7 & -\\
            \bottomrule[1.5pt]
        \end{tabular}
        \label{tab:my_label}
    \end{table}
    
    ~\cref{tab:my_label} summarizes the quantitative comparison between the given contact angles and the numerical obtained ones. It reveals that the current wetting boundary treatments are able to obtain satisfactory results for the entire range of contact angles from $10\degree$ to $160\degree$ for 2D simulations, with the maximum absolute errors generally falling below $1\degree$. A contact angle above 150$\degree$ indicates super-hydrophobic wetting properties. Numerically modeling the contact angle within such wetting regions poses challenges due to significant interface deformation, which can potentially lead to numerical instability. However, it is shown that the predicted values are overall consistent with the prescribed one, even when for the case with contact angle $\theta=170\degree$, the absolute error is also less than $2\degree$, which proves the good performance of the present scheme.
    
    In the 3D testing cases, the initial conditions and parameter settings are basically consistent with those of the 2D scenario. The simulations are performed in $Nx \times Ny \times Nz = 256 \times 256 \times 128$ domain.
    The initial distribution profile of the order parameter is given by
    \begin{equation}
        \phi(x,y,z)=0.5+0.5\tanh \frac{2[R-\sqrt{(x-x_0)^2+(y-y_0)^2+(z-z_0)^2}]}{W}.
    \end{equation}
    where $x_0=0.5Nx$, $y_0=0.5Ny$ and $z_0=0.25Nz$.~\cref{3Dflat} depicts the 3D equilibrium shapes of the droplet at different wetting angles, and~\cref{tab:my_label} presents the corresponding computed contact angles. From~\cref{tab:my_label}, we can find that the present scheme is able to obtain satisfactory results for the entire range of contact angles from 10$\degree$ to 140$\degree$, of which the maximum absolute errors are almost less than 3$\degree$. However, for the super-hydrophobic wetting condition, the present boundary treatment work with poor performance, especially for the cases with $\theta=160\degree$ and $170\degree$, where the droplets could be detached from the solid substrate, leading to a large prediction error. Compared to the 2D results, the 3D cases exhibit poorer performance in predicting super-hydrophobic wetting. This can be attributed to the smaller contact area in 3D cases compared to two. In the 2D cases, the contact line essentially acts as a contact surface of infinite length in the third dimension. This is considerably larger than the contact area in the 3D cases, which gradually reduces to a contact point as the wetting angle increases.  Its inability to provide sufficient adhesion finally leads to the detachment of the droplet~\cite{liang2019lattice}.
    
    \subsection{Droplet spreading on the inclined ideal surface}
    \begin{figure}[ht]
         \centering
        {\includegraphics[width=0.4\textwidth]{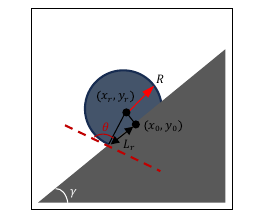}}~~
         \caption{Schematic illustration for the wetting of a droplet on an idea inclined surface, $\theta$ is the contact angle, $\gamma$ represents the inclined angle of the solid boundary.}
         \label{inclineshiyi}
    \end{figure}
    To demonstrate the applicability of the present wetting boundary treatment for problems with more general geometries, testing cases, where  droplets spreading on the inclined ideal surface with different inclination angles are performed. The 2D configuration of the problem is shown in~\cref{inclineshiyi}. As shown in this figure, an inclined solid is placed in $Nx \times Ny = 256 \times 256$ rectangular domain for 2D simulations, with its slanted edge represented by 
    $y=\tan{(\gamma)}x$, where $\gamma$ is the inclination angle, and it is selected as $\gamma=\arctan{(1.0)}$, $\arctan{(0.5)}$ and $\arctan{(0.25)}$ to assess the accuracy of the proposed method. Initially, semicircular droplet with the radius $R = 50$ is placed on the inclined surface, and the order parameter can be described by~\cref{initialphi}, with $(x_0, y_0)=(0.5N_x,\tan{(\gamma)}0.5N_x)$. Some parameters in the testing cases are set as $\rho_l=10.0$, $\rho_g=1.0$, $\nu_l=\nu_g=0.1$, $\sigma=0.005$ and $M=0.01$, and the wetting boundary treatments are adopted for the fluid-solid interface. For 2D cases, the interface thickness is set as $W=4$.
    
    \begin{figure}[ht]
         \centering
        {\includegraphics[width=0.9\textwidth]{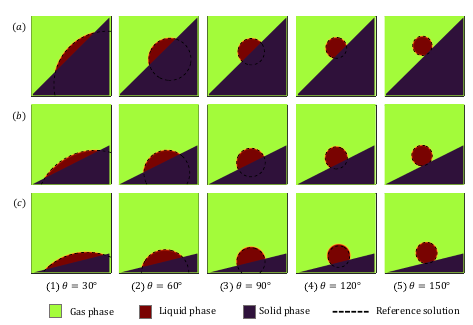}}~~
         \caption{The predicted droplet equilibrium shapes on the inclined surface in 2D using the present boundary treatment with wide range of prescribed contact angles and different inclined angles $\gamma$, (a) $\gamma=\arctan(1.0)$, (b) $\gamma=\arctan(0.5)$, (c) $\gamma=\arctan(0.25)$. }
         \label{2Dincline}
    \end{figure}
\begin{figure}[ht]
         \centering
        {\includegraphics[width=0.65\textwidth]{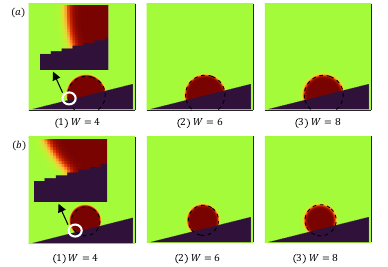}}~~
         \caption{The predicted droplet equilibrium shapes in 2D using the wetting boundary treatment with different width of phase interface $W$; contact angle (a) $\theta=90\degree$ and (b) $\theta=120\degree$. }
         \label{2Dincline2}
    \end{figure}

    ~\cref{2Dincline} shows the simulation results with different wall inclination angles for different contact angles from $\theta=30\degree$ to $150\degree$. 
    For $\gamma=\arctan(1)$ and $\arctan(0.5)$, it can be seen that the numerical results agree well with the reference solutions which is a set of circles with specific contact angles on the wetting surface obtained from geometric relationships. However, for the cases with $\gamma=\arctan{(0.25)}$, there is noticeable discrepancies between the numerical results and the reference solution, when $\theta=90\degree$ and $120\degree$. This discrepancy arises from the stair-stepped grid approximation for the inclined surface. A smaller inclination angle of the wall results in an elongated horizontal platform (as shown in~\cref{2Dincline2}). In some certain scenarios, the tri-phase contact point may be located on this platform, leading to a deviation in the prediction of the contact angle.
    \begin{figure}[ht]
         \centering
        {\includegraphics[width=0.9\textwidth]{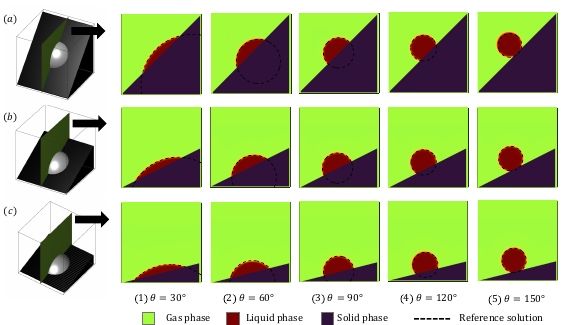}}~~
         \caption{The predicted droplet equilibrium shapes on the inclined surface in 3D using the present boundary treatment with wide range of prescribed contact angles and different inclined angles $\gamma$ (a) $\gamma=\arctan(1.0)$, (b) $\gamma=\arctan(0.5)$, (c) $\gamma=\arctan(0.25)$.}
         \label{3Dincline}
    \end{figure}
    To validate this perspective, an increased interface thickness 
    $W=6$ and 8 were adopted to ensure that the three-phase contact region encompasses a broader extent of solid grid points, thereby enhancing the accuracy of the stair-stepped approximation.
    ~\cref{2Dincline2} shows the numerical results for wetting angles of $90\degree$ and $120\degree$ with interface thicknesses $W=6$ and 8. As observed, the alignment between the calculated and reference results improves with increasing interface thickness. These numerical findings substantiate the accuracy of the proposed scheme in 2D scenarios.
    
    For the 3D validation, the simulations are performed in $Nx \times Ny \times Nz = 256 \times 256 \times 256$ domain.
    The slanted edge of the solid can be represented by $z=\tan{(\gamma)}x$. A droplet, shaped as a semicircle with a radius of $R=50$, is positioned on the tilted surface with its center located at $(x_0, y_0, z_0)=(0.5N_x, 0.5N_y, \tan{(\gamma)}0.5N_x)$.
    Aside from setting the interface thickness to $W=6$, all other parameters remain consistent with the two-dimensional setup. ~\cref{3Dincline} shows the a comparison between the numerical solution and the reference solution. It can be observed that for various inclinations of the inclined surface, the wetting boundary treatment proposed in this study accurately predicts the wetting angle. The numerical solution aligns closely with the reference solution, demonstrating the capability of the proposed method in addressing three-dimensional wetting problems.
    
    \subsection{Droplet spreading on the cylindrical and sphere surface}
    In the above subsections, the performance of the proposed boundary treatment on flat walls has been proved, including the both scenarios where the physical boundary aligns with the lattice link or not. In this subsection, the accuracy of the present model to enforce a designated contact angle on a curved boundary will be validated using the equilibrium configuration of a stationary droplet on a 2D circular surface and a 3D sphere.
    
    For 2D simulations, the A circular solid with a radius of $R_s=60$ and center located at $(x_s,y_s)=(0.5N_x, 0.5N_y-50)$ is placed within a rectangular area of size $N_x\times N_y=256\times 256$. Initially, a droplet with a radius of $R_i=50$ is placed on the surface of solid with the initial order parameter can be expressed as,
    \begin{equation}\label{initialphi_cir}
        \phi(x,y)=0.5+0.5\tanh \frac{2[R_i-\sqrt{(x-x_0)^2+(y-y_0)^2}]}{W},
    \end{equation}
    where $(x_0,y_0)=(0.5N_x, 0.5N_y)$ is the initial center location of the droplet.
    \begin{figure}[ht]
         \centering
        {\includegraphics[width=0.25\textwidth]{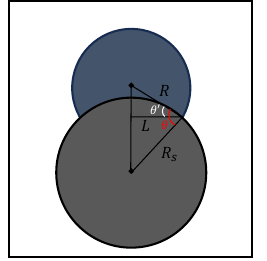}}~~
         \caption{Schematic illustration for the wetting of a droplet on a cylindrical solid, $\theta$ is the contact angle.}
         \label{cirshiyi}
    \end{figure}
    At equilibrium, the free energy of the system minimizes, suggesting an inherent tendency for the droplet to assume a circular form. ~\cref{cirshiyi} illustrates this equilibrium configuration of a droplet resting on a circular surface, and the reference solution can be expressed as
    \begin{equation}\label{initialphi_cirref}
        \phi_f(x,y)=0.5+0.5\tanh \frac{2[R-\sqrt{(x-x_e)^2+(y-y_e)^2}]}{W},
    \end{equation}
    where $R=L/\cos \theta'$ is the radius of the equilibrium droplet. $L$ can be obtained from the numerical solution and $\theta'=\theta-\theta^{*}$ with $\theta^{*}=\arccos(L/R_s)$. $(x_e,y_e)$ is the center location of the equilibrium droplet, where $x_e=x_s$, and $y_e=y_s+\sqrt{R^2+R_s^2-2RR_s\cos{\theta}}$.
    Parameters in simulations in the these testing cases are set as $\rho_l=10.0$, $\rho_g=1.0$, $\nu_l=\nu_g=0.1$, $\sigma=0.005$, $M=0.01$ and $W=6$.
    
    \begin{figure}[ht]
         \centering
        {\includegraphics[width=0.9\textwidth]{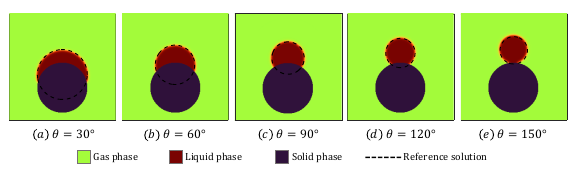}}~~
         \caption{The predicted droplet equilibrium shapes on the cylindrical solid in 2D using the present boundary treatment with wide range of prescribed contact angles.}
         \label{2Dcir}
    \end{figure}
    ~\cref{2Dcir} shows the comparison between the numerical results and the reference solutions for contact angles ranging from $30\degree$ to $150\degree$. As we can see, the numerical results agree well with the reference solutions, which proves the good performance of the proposed boundary treatment for 2D curved wall.
    
    \begin{figure}[ht]
         \centering
        {\includegraphics[width=0.9\textwidth]{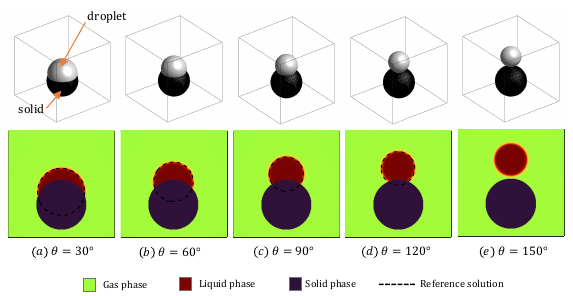}}~~
         \caption{The predicted droplet equilibrium shapes on a 3D solid sphere using the present boundary treatment with wide range of prescribed contact angles.The first row shows the 3D wetting morphology of the droplet, and the second row provides the distribution of order parameters in the central cross-section, along with a comparison to the reference solution. }
         \label{3Dcir}
    \end{figure}
    For 3D testing cases, the computational domain is $N_x\times N_y\times N_z=256\times 256\times 256$, and the initial distribution of the order parameter can be expressed as
    \begin{equation}
        \phi(x,y,z)=0.5+0.5\tanh \frac{2[R_i-\sqrt{(x-x_0)^2+(y-y_0)^2+(z-z_0)^2}]}{W}.
    \end{equation}
    All the other simulation and geometrical parameters are set as same as those in 2D cases.
    ~\cref{3Dcir} displays the 3D wetting morphology of the droplet at equilibrium, along with the phase field distribution on the cross-section at $y=0.5N_y$ and its corresponding reference solution. From the figure, it can be observed that when the contact angle is less than $150\degree$, the numerical results agree well with the reference solutions. However, for the case with the designated contact angle $\theta=150\degree$, the present boundary treatment work with poor performance. This serious deviation may be attributed to the fact that the droplet achieves a small wetting area on the sphere, as a result, the droplet could be detached from the solid substrate, which is similar with the testing cases on the flat wetting boundary, and it has been discussed in the above subsection. 
    
    \section{Conclusions}\label{sec4}
    In the present work, a simplified method is proposed to implement the free-energy based wetting boundary condition based on the phase-field LBE method. The two phase flow behavior is described by a well-balanced LBE model, and the wetting phenomena is governed by the surface free energy. Taking the example of the surface free energy in a cubic form, the proposed approach incorporates a portion of the surface free energy into the chemical potential. Unlike previous methods that traditionally treated it as a boundary condition, the proposed approach only requires handling the boundary condition with zero gradients of the order parameter in the normal direction on the solid nodes, which can be approximated by the average value of the sorrounding nodes. This approach significantly simplifies the implementation complexity of the wetting boundary condition. Several benchmark testing cases including the 2D and 3D droplet spreading processes on the flat, inclined and curved ideal walls were carried out to validate the accuracy of the proposed scheme. The results indicate the good ability and satisfactory accuracy of the proposed schemes to simulate wetting phenomena on curved boundaries. The boundary treatment proposed in this paper provides a simple and effective tool for the numerical simulation of multiphase flow in porous media .
    
    \section*{Declaration of Competing Interest}
    No Conflict of Interest declared.
    \section*{Data Availability Statements}
    The data that support the findings of this study are available from the corresponding author upon reasonable request.
    \section*{Acknowledgments}
    L.J. gratefully acknowledges insightful discussions with Prof. H. Liang in Hangzhou Dianzi University.
    L.J., S.S and B.Y would like to express appreciation to King Abdullah University of Science and Technology (KAUST) for the support through the grants BAS/1/1351-01, URF/1/5028-01 and BAS/1/1423-01-01.
    This work was also supported by the National Natural Science Foundation of China (No. 51836003 ) and the Interdiciplinary Research Program of Hust (2023JCYJ002).
\appendix

\section{Numerical verification of the irrelevance of $\chi$ to the results}
\label{appA}
In this section, we want to numerically prove that it is equivalent to treat the wall free energy at the boundary condition or to embody it in the chemical potential. Thus three different groups of chemical potential $\mu_{\phi}$ and $\chi$ are chosen to be the testing cases, which are listed in~\cref{table2}.
\begin{table}[ht]
    \centering
    \caption{Selection of chemical potential $\mu_{\phi}$ and $\chi$ in different cases.}
    \begin{tabular}{ccc}
    \toprule[1.5pt]
        \makebox[0.12\textwidth][c]{ } & \makebox[0.2\textwidth][c]{\makecell{$\chi$}} & \makebox[0.2\textwidth][c]{\makecell{chemical potential $\mu_{\phi}$}}\\
        \midrule[1pt]
        case 1 &  $-\sqrt{{2\beta}/{\kappa}}\cos\theta(\phi-\phi^2)$ &  $\frac{d \psi}{d{\phi}}-\kappa\nabla^2{\phi}$\\
        case 2 &  0&  $\frac{d \psi}{d{\phi}}-\kappa\nabla^2{\phi}-a_v\left[\sqrt{2\kappa\beta}\cos\theta({\phi}-{\phi}^2)\right]$\\
        case 3 &  $\sqrt{{2\beta}/{\kappa}}\cos\theta(\phi-\phi^2)$& $\frac{d \psi}{d{\phi}}-\kappa\nabla^2{\phi}-2a_v\left[\sqrt{2\kappa\beta}\cos\theta({\phi}-{\phi}^2)\right]$\\
        \bottomrule[1.5pt]
    \end{tabular}
    \label{table2}
\end{table}

\begin{figure}[ht]
    \centering
   {\includegraphics[width=0.9\textwidth]{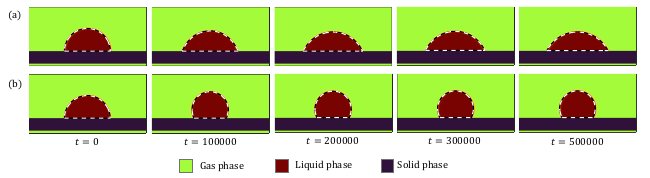}}~~
    \caption{Time evolution of the droplet spreading shapes with different contact angles, (a) $\theta=60\degree$, (b) $\theta=120\degree$. The results of case 1 with $\chi=-\sqrt{{2\beta}/{\kappa}}\cos\theta(\phi-\phi^2)$ are shown with color figure; The contours with $\phi=0.5$ of case 2 and case 3 are marked with black line and white dash line respectively. }
    \label{compare}
\end{figure}

In order to facilitate the calculation of cases 1 and 3, we selected the spreading process of droplets on an ideal horizontal plane as the test. The physical design and numerical parameters of the test are all consistent with that in~\cref{sec3.1}.
~\cref{compare} shows the time evolution of the droplet spreading shapes obtained by different cases. It can be clearly seen that at different contact angles, the results obtained at different times by different combinations of $\mu_{\phi}$ and $\chi$ are consistent, which proves the the irrelevance of $\chi$ to the wetting boundary treatment.
\label{appA}.

\bibliographystyle{elsarticle-num} 
\bibliography{main}

\begin{thebibliography}{10}
\expandafter\ifx\csname url\endcsname\relax
  \def\url#1{\texttt{#1}}\fi
\expandafter\ifx\csname urlprefix\endcsname\relax\def\urlprefix{URL }\fi
\expandafter\ifx\csname href\endcsname\relax
  \def\href#1#2{#2} \def\path#1{#1}\fi

\bibitem{alvarado2010enhanced}
V.~Alvarado, E.~Manrique, Enhanced oil recovery: an update review, Energies
  3~(9) (2010) 1529--1575.

\bibitem{yan2018multi}
B.~Yan, L.~Mi, Y.~Wang, H.~Tang, C.~An, J.~E. Killough, Multi-porosity
  multi-physics compositional simulation for gas storage and transport in
  highly heterogeneous shales, Journal of Petroleum Science and Engineering 160
  (2018) 498--509.

\bibitem{zhang2014mechanisms}
D.~Zhang, J.~Song, Mechanisms for geological carbon sequestration, Procedia
  IUTAm 10 (2014) 319--327.

\bibitem{depaolo2013geochemistry}
D.~J. DePaolo, D.~R. Cole, Geochemistry of geologic carbon sequestration: an
  overview, Reviews in Mineralogy and Geochemistry 77~(1) (2013) 1--14.

\bibitem{tariq2023spatial}
Z.~Tariq, E.~U. Yildirim, M.~Gudala, B.~Yan, S.~Sun, H.~Hoteit,
  Spatial--temporal prediction of minerals dissolution and precipitation using
  deep learning techniques: An implication to geological carbon sequestration,
  Fuel 341 (2023) 127677.

\bibitem{yan2023robust}
B.~Yan, M.~Gudala, S.~Sun, Robust optimization of geothermal recovery based on
  a generalized thermal decline model and deep learning, Energy Conversion and
  Management 286 (2023) 117033.

\bibitem{huang2023compositional}
T.~Huang, G.~J. Moridis, T.~A. Blasingame, A.~M. Abdulkader, B.~Yan,
  Compositional reservoir simulation of underground hydrogen storage in
  depleted gas reservoirs, International Journal of Hydrogen Energy 48 (2023)
  36035--36050.

\bibitem{song2021pore}
R.~Song, S.~Sun, J.~Liu, C.~Yang, Pore scale modeling on dissociation and
  transportation of methane hydrate in porous sediments, Energy 237 (2021)
  121630.

\bibitem{de1985wetting}
D.~Gennes, G.~Pierre, Wetting: statics and dynamics, Reviews of modern physics
  57~(3) (1985) 827--863.

\bibitem{cox1986dynamics1}
R.~Cox, The dynamics of the spreading of liquids on a solid surface. part 1.
  viscous flow, Journal of fluid mechanics 168 (1986) 169--194.

\bibitem{cox1986dynamics2}
R.~Cox, The dynamics of the spreading of liquids on a solid surface. part 2.
  surfactants, Journal of Fluid mechanics 168 (1986) 195--220.

\bibitem{zhang2008theoretical}
H.~Zhang, R.~Wang, D.~T. Liang, J.~H. Tay, Theoretical and experimental studies
  of membrane wetting in the membrane gas--liquid contacting process for co2
  absorption, Journal of Membrane Science 308~(1-2) (2008) 162--170.

\bibitem{wei2023experimental}
J.~Wei, W.~Jiang, L.~Si, X.~Xu, Z.~Wen, Experimental study of wetting effect of
  surfactant based on dynamic wetting process and impedance response of coal,
  Environmental Science and Pollution Research 30~(2) (2023) 4278--4292.

\bibitem{mukherjee2007numerical}
A.~Mukherjee, S.~G. Kandlikar, Numerical study of single bubbles with dynamic
  contact angle during nucleate pool boiling, International Journal of Heat and
  Mass Transfer 50~(1-2) (2007) 127--138.

\bibitem{chaudhary2014freezing}
G.~Chaudhary, R.~Li, Freezing of water droplets on solid surfaces: An
  experimental and numerical study, Experimental Thermal and Fluid Science 57
  (2014) 86--93.

\bibitem{guo2013lattice}
Z.~Guo, C.~Shu, Lattice Boltzmann method and its application in engineering,
  Vol.~3, World Scientific, 2013.

\bibitem{liu2016lattice}
H.~Liu, L.~Wu, Y.~Ba, G.~Xi, Y.~Zhang, A lattice boltzmann method for
  axisymmetric multicomponent flows with high viscosity ratio, Journal of
  Computational Physics 327 (2016) 873--893.

\bibitem{ba2016multiple}
Y.~Ba, H.~Liu, Q.~Li, Q.~Kang, J.~Sun, Multiple-relaxation-time color-gradient
  lattice boltzmann model for simulating two-phase flows with high density
  ratio, Physical Review E 94~(2) (2016) 023310.

\bibitem{akai2018wetting}
T.~Akai, B.~Bijeljic, M.~J. Blunt, Wetting boundary condition for the
  color-gradient lattice boltzmann method: Validation with analytical and
  experimental data, Advances in Water Resources 116 (2018) 56--66.

\bibitem{chen2014critical}
L.~Chen, Q.~Kang, Y.~Mu, Y.-L. He, W.-Q. Tao, A critical review of the
  pseudopotential multiphase lattice boltzmann model: Methods and applications,
  International journal of heat and mass transfer 76 (2014) 210--236.

\bibitem{li2013lattice}
Q.~Li, K.~Luo, X.~Li, Lattice boltzmann modeling of multiphase flows at large
  density ratio with an improved pseudopotential model, Physical Review E
  87~(5) (2013) 053301.

\bibitem{li2007symmetric}
Q.~Li, A.~Wagner, Symmetric free-energy-based multicomponent lattice boltzmann
  method, Physical Review E 76~(3) (2007) 036701.

\bibitem{soomro2023fugacity}
M.~Soomro, L.~F. Ayala, C.~Peng, O.~M. Ayala, Fugacity-based lattice boltzmann
  method for multicomponent multiphase systems, Physical Review E 107~(1)
  (2023) 015304.

\bibitem{guo2021well}
Z.~Guo, Well-balanced lattice boltzmann model for two-phase systems, Physics of
  Fluids 33~(3) (2021).

\bibitem{fakhari2010phase}
A.~Fakhari, M.~H. Rahimian, Phase-field modeling by the method of lattice
  boltzmann equations, Physical Review E 81~(3) (2010) 036707.

\bibitem{shu2013direct}
S.~Shu, N.~Yang, Direct numerical simulation of bubble dynamics using
  phase-field model and lattice boltzmann method, Industrial \& Engineering
  Chemistry Research 52~(33) (2013) 11391--11403.

\bibitem{yue2022improved}
L.~Yue, Z.~Chai, H.~Wang, B.~Shi, Improved phase-field-based lattice boltzmann
  method for thermocapillary flow, Physical Review E 105~(1) (2022) 015314.

\bibitem{zhang2023simplified}
S.~Zhang, J.~Tang, H.~Wu, Simplified wetting boundary scheme in phase-field
  lattice boltzmann model for wetting phenomena on curved boundaries, Physical
  Review E 108~(2) (2023) 025303.

\bibitem{martys1996simulation}
N.~S. Martys, H.~Chen, Simulation of multicomponent fluids in complex
  three-dimensional geometries by the lattice boltzmann method, Physical review
  E 53~(1) (1996) 743.

\bibitem{iwahara2003liquid}
D.~Iwahara, H.~Shinto, M.~Miyahara, K.~Higashitani, Liquid drops on homogeneous
  and chemically heterogeneous surfaces: A two-dimensional lattice boltzmann
  study, Langmuir 19~(21) (2003) 9086--9093.

\bibitem{ding2007wetting}
H.~Ding, P.~D. Spelt, Wetting condition in diffuse interface simulations of
  contact line motion, Physical Review E 75~(4) (2007) 046708.

\bibitem{wang2013scheme}
L.~Wang, H.~B. Huang, X.~Y. Lu, Scheme for contact angle and its hysteresis in
  a multiphase lattice boltzmann method, Physical Review E 87~(1) (2013)
  013301.

\bibitem{liu2015diffuse}
H.~R. Liu, H.~Ding, A diffuse-interface immersed-boundary method for
  two-dimensional simulation of flows with moving contact lines on curved
  substrates, Journal of Computational Physics 294 (2015) 484--502.

\bibitem{huang2018alternative}
J.~Huang, J.~Wu, H.~Huang, An alternative method to implement contact angle
  boundary condition and its application in hybrid lattice-boltzmann
  finite-difference simulations of two-phase flows with immersed surfaces, The
  European Physical Journal E 41 (2018) 1--18.

\bibitem{huang2022simplified}
J.~Huang, L.~Zhang, Simplified method for wetting on curved boundaries in
  conservative phase-field lattice-boltzmann simulation of two-phase flows with
  large density ratios, Physics of Fluids 34~(8) (2022).

\bibitem{briant2002lattice}
A.~Briant, Lattice boltzmann simulations of contact line motion in a liquid-gas
  system, Philosophical Transactions of the Royal Society of London. Series A:
  Mathematical, Physical and Engineering Sciences 360~(1792) (2002) 485--495.

\bibitem{briant2004lattice}
A.~Briant, J.~Yeomans, Lattice boltzmann simulations of contact line motion.
  ii. binary fluids, Physical Review E 69~(3) (2004) 031603.

\bibitem{lee2008wall}
T.~Lee, L.~Liu, Wall boundary conditions in the lattice boltzmann equation
  method for nonideal gases, Physical Review E 78~(1) (2008) 017702.

\bibitem{2009WALL}
L.~Liu, T.~Lee, Wall free energy based polynomial boundary conditions for
  non-ideal gas lattice boltzmann equation, International Journal of Modern
  Physics C 20~(11) (2009) 1749--1768.

\bibitem{qian2003molecular}
T.~Qian, X.-P. Wang, P.~Sheng, Molecular scale contact line hydrodynamics of
  immiscible flows, Physical Review E 68~(1) (2003) 016306.

\bibitem{villanueva2006some}
W.~Villanueva, G.~Amberg, Some generic capillary-driven flows, International
  Journal of Multiphase Flow 32~(9) (2006) 1072--1086.

\bibitem{khatavkar2007capillary}
V.~Khatavkar, P.~Anderson, H.~Meijer, Capillary spreading of a droplet in the
  partially wetting regime using a diffuse-interface model, Journal of Fluid
  Mechanics 572 (2007) 367--387.

\bibitem{CONNINGTON2013601}
K.~Connington, T.~Lee, Lattice boltzmann simulations of forced wetting
  transitions of drops on superhydrophobic surfaces, Journal of Computational
  Physics 250 (2013) 601--615.

\bibitem{CONNINGTON2015453}
K.~W. Connington, T.~Lee, J.~F. Morris, Interaction of fluid interfaces with
  immersed solid particles using the lattice boltzmann method for
  liquid–gas–particle systems, Journal of Computational Physics 283 (2015)
  453--477.

\bibitem{FAKHARI2017620}
A.~Fakhari, D.~Bolster, Diffuse interface modeling of three-phase contact line
  dynamics on curved boundaries: A lattice boltzmann model for large density
  and viscosity ratios, Journal of Computational Physics 334 (2017) 620--638.

\bibitem{FAKHARI2018119}
A.~Fakhari, Y.~Li, D.~Bolster, K.~T. Christensen, A phase-field lattice
  boltzmann model for simulating multiphase flows in porous media: Application
  and comparison to experiments of co2 sequestration at pore scale, Advances in
  Water Resources 114 (2018) 119--134.

\bibitem{zhang2019fractional}
C.~Zhang, Z.~Guo, Y.~Li, A fractional step lattice boltzmann model for
  two-phase flow with large density differences, International Journal of Heat
  and Mass Transfer 138 (2019) 1128--1141.

\bibitem{liang2018phase}
H.~Liang, J.~Xu, J.~Chen, H.~Wang, Z.~Chai, B.~Shi, Phase-field-based lattice
  boltzmann modeling of large-density-ratio two-phase flows, Physical Review E
  97~(3) (2018) 033309.

\bibitem{guo2002discrete}
Z.~Guo, C.~Zheng, B.~Shi, Discrete lattice effects on the forcing term in the
  lattice boltzmann method, Physical review E 65~(4) (2002) 046308.

\bibitem{2311.10827}
L.~Ju, P.~Liu, B.~Yan, J.~Bao, S.~Sun, Z.~Guo, A well-balanced lattice
  boltzmann model for binary fluids based on the incompressible phase-field
  theory (2023).
\newblock \href {http://arxiv.org/abs/arXiv:2311.10827}
  {\path{arXiv:arXiv:2311.10827}}.

\bibitem{liang2014phase}
H.~Liang, B.~Shi, Z.~Guo, Z.~Chai, Phase-field-based multiple-relaxation-time
  lattice boltzmann model for incompressible multiphase flows, Physical Review
  E 89~(5) (2014) 053320.

\bibitem{moldover1980interface}
M.~R. Moldover, J.~W. Cahn, An interface phase transition: complete to partial
  wetting, Science 207~(4435) (1980) 1073--1075.

\bibitem{liang2019lattice}
H.~Liang, H.~Liu, Z.~Chai, B.~Shi, Lattice boltzmann method for contact-line
  motion of binary fluids with high density ratio, Physical Review E 99~(6)
  (2019) 063306.

\bibitem{yan2007lattice}
Y.~Yan, Y.~Zu, A lattice boltzmann method for incompressible two-phase flows on
  partial wetting surface with large density ratio, Journal of Computational
  Physics 227~(1) (2007) 763--775.

\bibitem{young1805iii}
T.~Young, Iii. an essay on the cohesion of fluids, Philosophical transactions
  of the royal society of London~(95) (1805) 65--87.

\bibitem{li2009solving}
X.~Li, J.~Lowengrub, A.~R{\"a}tz, A.~Voigt, Solving pdes in complex geometries:
  a diffuse domain approach, Communications in mathematical sciences 7~(1)
  (2009) 81--107.

\bibitem{li2019implementation}
Q.~Li, Y.~Yu, K.~H. Luo, Implementation of contact angles in pseudopotential
  lattice boltzmann simulations with curved boundaries, Physical review E
  100~(5) (2019) 053313.

\bibitem{whitaker1998method}
S.~Whitaker, The method of volume averaging, Vol.~13, Springer Science \&
  Business Media, 1998.

\bibitem{soulaine2017mineral}
C.~Soulaine, S.~Roman, A.~Kovscek, H.~A. Tchelepi, Mineral dissolution and
  wormholing from a pore-scale perspective, Journal of Fluid Mechanics 827
  (2017) 457--483.

\bibitem{liu2022diffuse}
X.~Liu, Z.~Chai, C.~Zhan, B.~Shi, W.~Zhang, A diffuse-domain phase-field
  lattice boltzmann method for two-phase flows in complex geometries,
  Multiscale Modeling \& Simulation 20~(4) (2022) 1411--1436.

\end{thebibliography}





\end{document}